\documentclass[11pt]{article}
\usepackage[letterpaper,margin=1in]{geometry}
\usepackage{amsmath,amssymb,amsfonts}
\usepackage{graphicx}
\usepackage{array,booktabs}
\usepackage{cite}
\usepackage{url}
\usepackage{lineno}
\usepackage{caption}
\usepackage{subfig}
\usepackage{microtype}
\usepackage[hidelinks]{hyperref}
\graphicspath{{Figures/}}
\title{Memristor-Based Pulse Width Modulation Circuit for Power Converters with Programmable Frequencies}
\author{Fanfu Wu$^{1}$, Xiaoyi Lei$^{1}$, and Yunting Liu$^{1,*}$\\[4pt]
\small $^{1}$Department of Electrical Engineering, The Pennsylvania State University, University Park, PA, USA\\
\small $^{*}$Corresponding author: ypl5778@psu.edu}
\date{}

\begin{document}
\maketitle

\begin{abstract}
Memristors can achieve up to a 1000× reduction in energy consumption in neuromorphic and in-memory computing, but their integration into power converter control remains incomplete. Existing implementations either convert memristor outputs into digital signals for DSP-based pulse-width modulation (PWM), reintroducing computational overhead, or use analog PWM circuits with limited programmability. Since PWM generation is the final essential stage of power converter control, a programmable memristor-based PWM circuit is needed to fully realize the energy-efficiency benefits of memristive computing. This paper presents a programmable memristor-based PWM circuit operating at approximately two-hundred kilohertz for switching power converters. The design combines digital-controller programmability with the low power consumption of analog PWM generators. Experimental validation using commercially available memristors achieves a programmable switching frequency from 144.7 kHz to 204.2 kHz. Compared with a commonly used DSP implementation in the power converter control, the proposed design achieves a 92\% reduction in power consumption.
\end{abstract}

\section*{Introduction}
Memristors are an emerging technology that offers significant potential for reducing computational power consumption by performing computation directly within memory arrays \cite{energyefficient,intro2,intro3,intro4,intro5,intro6,intro7}. In in-memory computing applications, memristor-based systems have been reported to reduce energy consumption by approximately 99.9\% compared with conventional computing architectures \cite{Power_in_memory}. Furthermore, energy savings of up to 308× have been demonstrated when compared with SRAM-based AI accelerators \cite{SRAM}. 
Besides the applications in AI, memristors also show great potential in analog circuit design\cite{mem_analog}. A programmable memristor-based Kalman filter and PD controller for robotic applications have been developed, demonstrating a 1546× reduction in energy consumption and a 1586× reduction in computation time\cite{PDcontrol}. Despite the energy-efficiency advantages of memristive computing, practical deployment in power converter control remains challenging because converter outputs must ultimately be translated into PWM signals.

\begin{figure}[t]
\centerline{\includegraphics[width=0.95\textwidth]{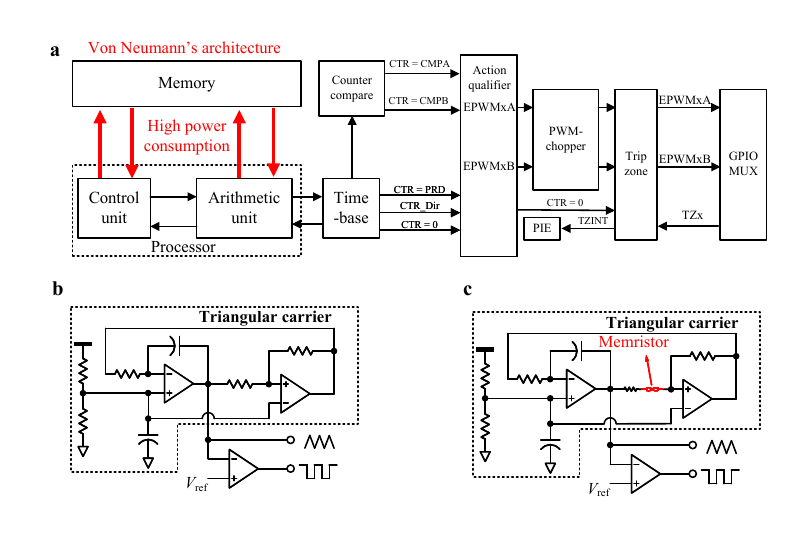}}
\caption{\textbf{Comparison of digital, conventional analog and
memristor-based PWM generation.}
\textbf{a} Representative PWM-generation path in a digital signal
processor.\cite{TIDSP}
\textbf{b} Conventional analog PWM generator.\cite{Analog_PWM}
\textbf{c} Proposed memristor-based PWM generator.}
\label{intro_circuit}
\end{figure}

Some potential solutions to integrate the emerging memristor-based computing with power converters include 1) leveraging digital PWM modules in digital signal processors (DSPs) through a memristor-to-DSP ADC interface and 2) directly interfacing the memristor with an analog PWM generator. DSPs are widely used in power converter control applications. In DSP-based systems, PWM signals are typically generated using time-based counters and comparators \cite{TIDSP}, where the counter value is compared with reference registers (CMPA, CMPB) to determine the duty cycle, as illustrated in Fig. \ref{intro_circuit}a \cite{TIDSP}. Although DSP-based PWM modules provide high programmability, they are not directly compatible with fully memristive computing platforms. The analog computation results generated by memristor crossbars must first be sensed, converted through ADC interfaces, stored or transferred in digital form, and then processed by the DSP to update the PWM reference registers \cite{Von}. This signal-conversion and memory-access process introduces additional interface complexity, latency, and power consumption. 
In contrast, analog PWM generators offer significantly lower power consumption. These circuits produce PWM signals by comparing a carrier signal generated by a triangular waveform with a reference voltage, as shown in Fig. \ref{intro_circuit}b \cite{Analog_PWM}. However, analog PWM modules are constrained by limited programmability, as changes in switching frequency or operating range typically require hardware modification. Modern power converters frequently require adaptive switching frequencies, reconfigurable modulation strategies, and parameter updates. Conventional analog PWM generators cannot readily support these functions. While analog PWM generators can be combined with memristor-based controllers without additional conversion or interface, their restricted flexibility limits their compatibility with programmable memristive systems.

Hassanein et al.~\cite{10HzPWM} were the first to demonstrate a memristor-based PWM generator. While simulation results validated the feasibility of the concept, the operating frequency was limited to 10 Hz, and the architecture required closely matched memristor dynamics, limiting its practical applicability. To address these limitations, we previously proposed a preliminary circuit that improved the practical implementation of memristor-based PWM generation and demonstrated operation at 20 kHz~\cite{20kHz}. However, the achievable switching frequency and programmability remained constrained by the low threshold voltage and stochastic switching characteristics of practical memristor devices. Building upon these prior developments, this paper presents a new memristor-based PWM architecture capable of operating at frequencies up to two hundred kilohertz while maintaining programmability, thereby extending memristor-based control to switching frequencies needed by modern power electronic converters. Such frequencies are necessary for modern SiC/GaN converters, where switching frequencies typically range from tens to hundreds of kilohertz. By employing the circuit topology shown in Fig. \ref{intro_circuit}c, the limitations imposed by the memristor characteristics are mitigated, enabling a wider programmable frequency range suitable for high-frequency power electronics applications. This work closes the missing interface between memristive computing and practical power converter control, enabling a fully programmable low-power control architecture.

\section*{Results}
\subsection*{Memristor characteristics}
\begin{figure}[htbp]
\centerline{\includegraphics[width=0.9\textwidth]{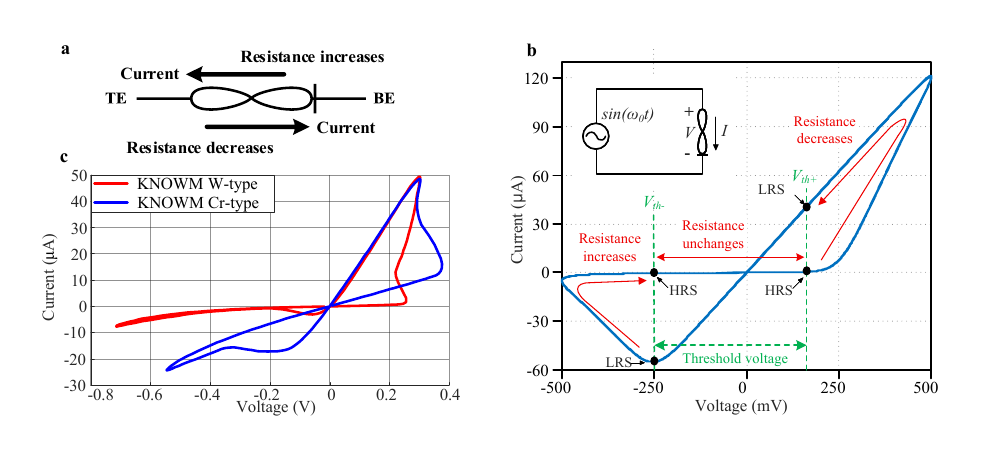}}
\vspace{-10pt}
\caption{\textbf{Operating characteristics and experimental
characterization of memristors.}
\textbf{a} Schematic illustration of the polarity-dependent resistance
change of a two-terminal memristor.
\textbf{b} Idealized current--voltage hysteresis of a memristor under sinusoidal excitation.
\textbf{c} Measured current–voltage hysteresis curves of commercial Knowm W-type and Cr-type memristors, averaged over 50 cycles.}
\label{Mem_fundamental}
\end{figure}

A memristor is a two-terminal device whose resistance depends on the history of the applied voltage or current. When an external voltage or current is removed, the device retains its resistance state, exhibiting nonvolatile behavior~\cite{fundamental_mem}. As illustrated in Fig. \ref{Mem_fundamental}a, it consists of a top electrode (TE) and a bottom electrode (BE). When current flows from the TE to the BE, the device resistance decreases; in the opposite direction, its resistance increases. Physically, the memristor exhibits a high-resistance state (HRS), corresponding to its maximum resistance, and a low-resistance state (LRS), corresponding to its minimum resistance. When a periodic voltage, such as a sinusoidal waveform, is applied across the memristor, its resistance state begins to change once the applied voltage exceeds the threshold voltage as shown in Fig~\ref{Mem_fundamental}b. This results in a unique hysteretic current–voltage characteristic, which distinguishes the device from a conventional resistor and confirms its memristive behavior. The mean hysteresis curves of the commercial tungsten (W)-type and chromium (Cr)-type memristors over 50 cycles are shown in Fig.~\ref{Mem_fundamental}c. Both devices exhibit pinched hysteresis loops, confirming their memristive behavior.

\subsection*{Memristor resistance programming} 
\begin{figure}[t]
    \centering
    \includegraphics[
        width=0.85\textwidth,
    ]{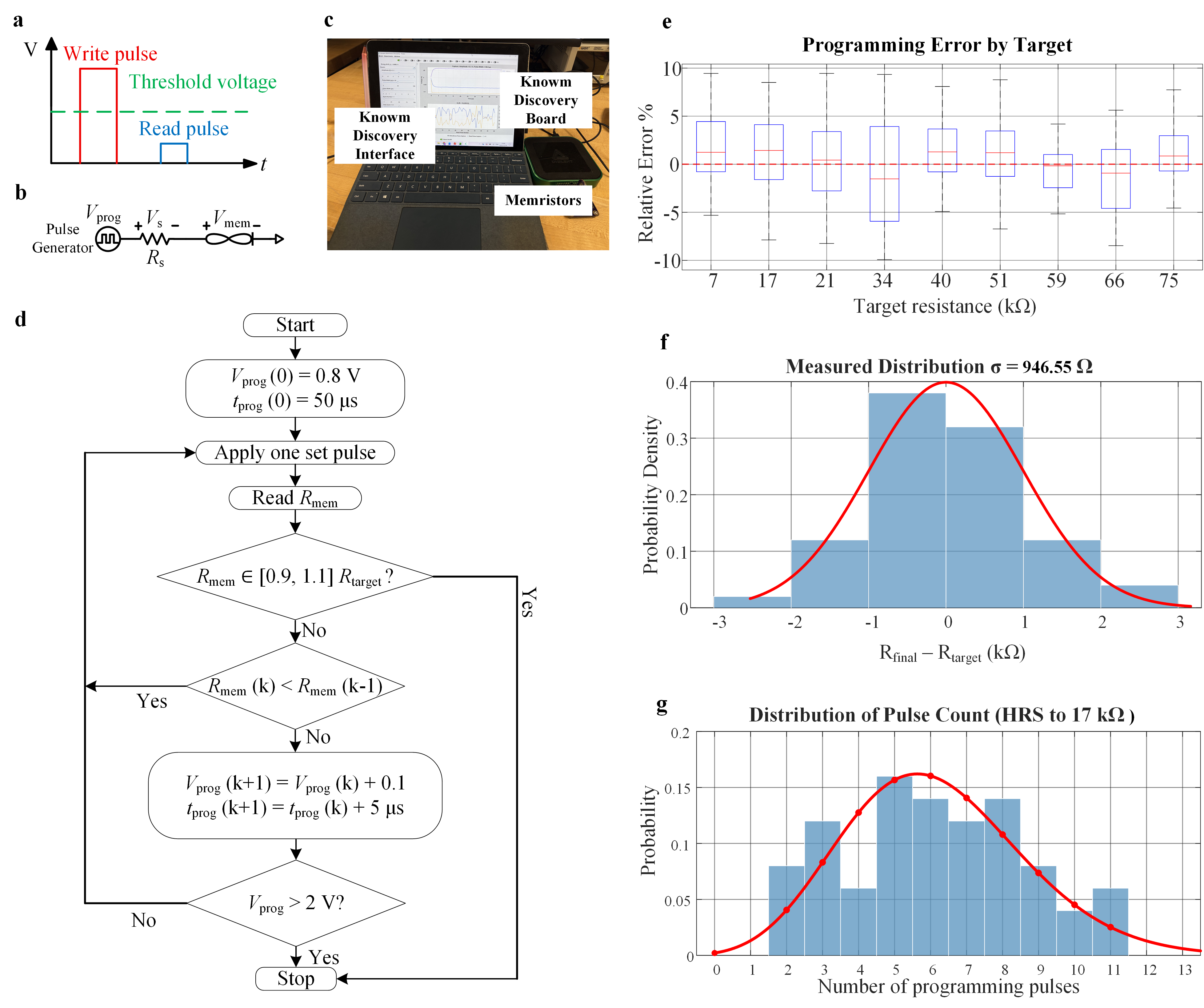}
    \caption{\textbf{Closed-loop programming and statistical characterization of the memristor resistance.}
    \textbf{a} Write--read pulses used for resistance programming.
    \textbf{b} Schematic of the programming circuit.
    \textbf{c} Experimental programming platform based on the Knowm Discovery board and its interface.
    \textbf{d} Closed-loop procedure for tuning the memristor from a high-resistance state to a target resistance.
    \textbf{e} Relative programming error for nine target resistance values.
    \textbf{f} Programming-error distribution for the 21-k$\Omega$ target, with a standard deviation of $\sigma=946.55~\Omega$.
    \textbf{g} Distribution of the number of set pulses required to program the memristor from the high-resistance state to 17~k$\Omega$.}
    \label{Mem_prog}
\end{figure}

The memristor can be programmed to many intermediate resistance states between HRS and LRS. To program the resistance state, voltage-pulse excitation is commonly used as shown in Fig \ref{Mem_prog}a. A write pulse with an amplitude larger than the memristor threshold voltage is applied across the memristor to change its resistance state. After the write operation, a read pulse with an amplitude lower than the threshold voltage is applied to measure the resistance without further disturbing the programmed state. The memristor resistance can then be calculated from the measured read current as
\begin{equation}
    R_{\mathrm{mem}} = \frac{V_{\mathrm{read}}}{I_{\mathrm{read}}}
    \label{eq:mem_resistance}
\end{equation}
where $V_{\mathrm{read}}$ is the voltage across the memristor during the read pulse and $I_{\mathrm{read}}$ is the measured read current.

In this work, the Knowm Discovery board and its software interface are used to program and monitor the resistance state of the memristor. The experimental setup of the Discovery board is shown in Fig.~\ref{Mem_prog}b. A series resistor, $R_s$, is connected in series with the memristor to limit the programming current. A pulse generator provides both the write and read pulses. With this voltage-divider configuration, the voltage applied across the memristor is
\begin{equation}
    V_{\mathrm{mem}} = V_{\mathrm{prog}}
    \frac{R_{\mathrm{mem}}}{R_{\mathrm{mem}} + R_s}
    \label{eq:mem_voltage_divider}
\end{equation}
where $V_{\mathrm{prog}}$ is the applied programming voltage from the pulse generator.

During the write operation, the voltage across the memristor should exceed the threshold voltage, $V_{\mathrm{th}}$, to change its resistance state:
\begin{equation}
    V_{\mathrm{mem}} = V_{\mathrm{write}}
    \frac{R_{\mathrm{mem}}}{R_{\mathrm{mem}} + R_s}
    > V_{\mathrm{th}} .
    \label{eq:write_condition_1}
\end{equation}
Therefore, the required write-pulse amplitude from the pulse generator should satisfy
\begin{equation}
    V_{\mathrm{write}} >
    V_{\mathrm{th}}
    \frac{R_{\mathrm{mem}} + R_s}{R_{\mathrm{mem}}} .
    \label{eq:write_condition_2}
\end{equation}

During the read operation, the voltage across the memristor should remain below the threshold voltage to avoid unintentionally changing its resistance state:
\begin{equation}
    V_{\mathrm{mem}} = V_{\mathrm{read}}
    \frac{R_{\mathrm{mem}}}{R_{\mathrm{mem}} + R_s}
    < V_{\mathrm{th}} .
    \label{eq:read_condition_1}
\end{equation}
Thus, the read-pulse amplitude should satisfy
\begin{equation}
    V_{\mathrm{read}} <
    V_{\mathrm{th}}
    \frac{R_{\mathrm{mem}} + R_s}{R_{\mathrm{mem}}} .
    \label{eq:read_condition_2}
\end{equation}

This condition ensures that the write pulse can program the memristor resistance, while the read pulse only measures the resistance without altering the programmed state.

The Knowm Discovery board shown in Fig.~\ref{Mem_prog}c uses a default series resistor of $10~\mathrm{k}\Omega$ and applies a read pulse with a magnitude of $0.1~\mathrm{V}$ at a frequency of $1~\mathrm{Hz}$. Although the Knowm memristor can exhibit a wide resistance range, from the kilo-ohm range to the mega-ohm range, the practical resistance range that can be accurately monitored is limited by the read-pulse amplitude, measurement noise, and the tolerance of the interface circuit.

Since the read-pulse amplitude is relatively small and the interface has a voltage tolerance of approximately $\pm 10~\mathrm{mV}$, the measured voltage can be significantly affected by noise when the voltage division ratio is too small or too large. Therefore, to accurately monitor the memristor resistance without being strongly affected by measurement noise, the preferred resistance range is selected as
\begin{equation}
    0.1R_s \leq R_{\mathrm{mem}} < 10R_s ,
    \label{eq:preferred_range_general}
\end{equation}
where $R_s$ is the series resistor and $R_{\mathrm{mem}}$ is the memristor resistance.

For the default Discovery board configuration with $R_s = 10~\mathrm{k}\Omega$, the preferred programmable resistance range becomes
\begin{equation}
    1~\mathrm{k}\Omega \leq R_{\mathrm{mem}} < 100~\mathrm{k}\Omega .
    \label{eq:preferred_range_default}
\end{equation}

If the memristor is intended to operate at higher or lower resistance states, the series resistor should be adjusted accordingly based on Equation~(\ref{eq:preferred_range_general}). Moreover, the programmable range can also be affected by the threshold voltage of the memristor. The typical threshold voltage of the Knowm Cr-type memristor is approximately 0.33 V~\cite{Knowm_datasheet}, which requires a low-amplitude read pulse. If a memristor has a higher threshold voltage, the read-pulse amplitude can be increased while still remaining below the threshold voltage. This can improve the noise immunity and expand the accurately measurable resistance range for the same series resistor.


To ensure that the proposed circuit can generate PWM signals at different desired frequencies, the memristor must be programmed accurately to the target resistance values. However, due to the stochastic switching behavior of currently available memristors, programming the device to an exact resistance value is challenging. Therefore, in this work, a programming tolerance of $\pm 10\%$ from the target resistance is considered acceptable.

A specific programming procedure using the Knowm Discovery interface is adopted, as shown in Fig.~\ref{Mem_prog}d. After each write pulse, the resistance state of the memristor is checked, and the subsequent write pulse is updated based on the measured resistance state. Several stable resistance states of a Knowm SDC Cr-type memristor are programmed, and each target resistance is tested 50 times to evaluate programming repeatability. The programming accuracy for different target resistance values is summarized in Fig.~\ref{Mem_prog}e and Table \ref{tab:programmed_resistance_summary}. 
All selected resistance states can be programmed within the $\pm 10\%$ tolerance window. The median programming error for each target resistance is close to zero, indicating that the closed-loop programming method does not introduce a strong systematic bias. However, the spread of the boxplots varies among different target resistance levels, showing that the programming repeatability is resistance-dependent. Therefore, Fig.~\ref{Mem_prog}e demonstrates that the Knowm SDC Cr-type memristor can provide multiple distinguishable resistance states suitable for frequency programming in the proposed PWM circuit.



The programming-error distribution for the 21~k$\Omega$ target resistance is shown in Fig.~\ref{Mem_prog}(b), where the programming error is defined as the difference between the measured resistance and the target resistance. 
The distribution is centered near zero, and the measured standard deviation is $\sigma = 946.55~\Omega$, corresponding to approximately $4.5\%$ of the target resistance. 
Since this value is smaller than the $\pm 10\%$ tolerance used in this work, the result confirms that the proposed programming procedure can repeatedly tune the memristor to a practically useful resistance range. Based on the repeated programming trials, the empirical probability of reaching the $\pm 10\%$ target window is 100\%.

The stochastic nature of the memristor programming process is further illustrated in Fig.~\ref{Mem_prog}g, which shows the distribution of the number of writing pulses required to tune the memristor from a high-resistance state $(>100~\mathrm{k}\Omega)$ to 17~k$\Omega$. The required pulse count varies among repeated trials, confirming that the resistance-update process is not fully deterministic. Approximately 82\% of the trials converge within approximately 3--9 programming pulses, while the full observed range is approximately 2--11 pulses. This result shows that applying a fixed number of programming pulses may not reliably produce the same final resistance state. 
In contrast, the adopted closed-loop programming procedure compensates for cycle-to-cycle variation by checking the resistance after each write pulse and iteratively updating the programming condition until the desired target range is reached.

Overall, these results indicate that although stochastic variation exists in the programming process, the Knowm SDC Cr-type memristor can still provide controllable and repeatable resistance states for frequency programming in the proposed PWM circuit. The boxplot results confirm that multiple target resistance states can be reached within the selected tolerance window, while the pulse-count distribution shows that closed-loop programming is necessary to compensate for the nondeterministic switching behavior of the memristor.

\begin{table}[htbp]
    \centering
    \caption{Statistical summary of programmed memristor resistance states}
    \label{tab:programmed_resistance_summary}
    \begin{tabular}{cccc}
        \hline
        $R_{\mathrm{target}}$ (k$\Omega$) &
        $R_{\mathrm{measured}}$ Mean (k$\Omega$) &
        Std. (k$\Omega$) &
        Variance (\%) \\
        \hline
        7  & 7.110  & 0.2827 & 3.9757 \\
        17 & 17.121 & 0.6921 & 4.0425 \\
        21 & 21.191 & 0.9466 & 4.4668 \\
        34 & 34.147 & 2.1003 & 6.2332 \\
        40 & 40.654 & 1.1276 & 2.7735 \\
        51 & 51.590 & 1.6601 & 3.2179 \\
        59 & 58.778 & 1.3290 & 2.2611 \\
        66 & 64.965 & 2.4139 & 3.7157 \\
        75 & 75.805 & 2.2241 & 2.9340 \\
        \hline
    \end{tabular}
\end{table}

\subsection*{Design principles of the memristor-based PWM circuit}
\begin{figure}[htbp]
    \centering
    \includegraphics[
        width=0.9\textwidth,
        ]{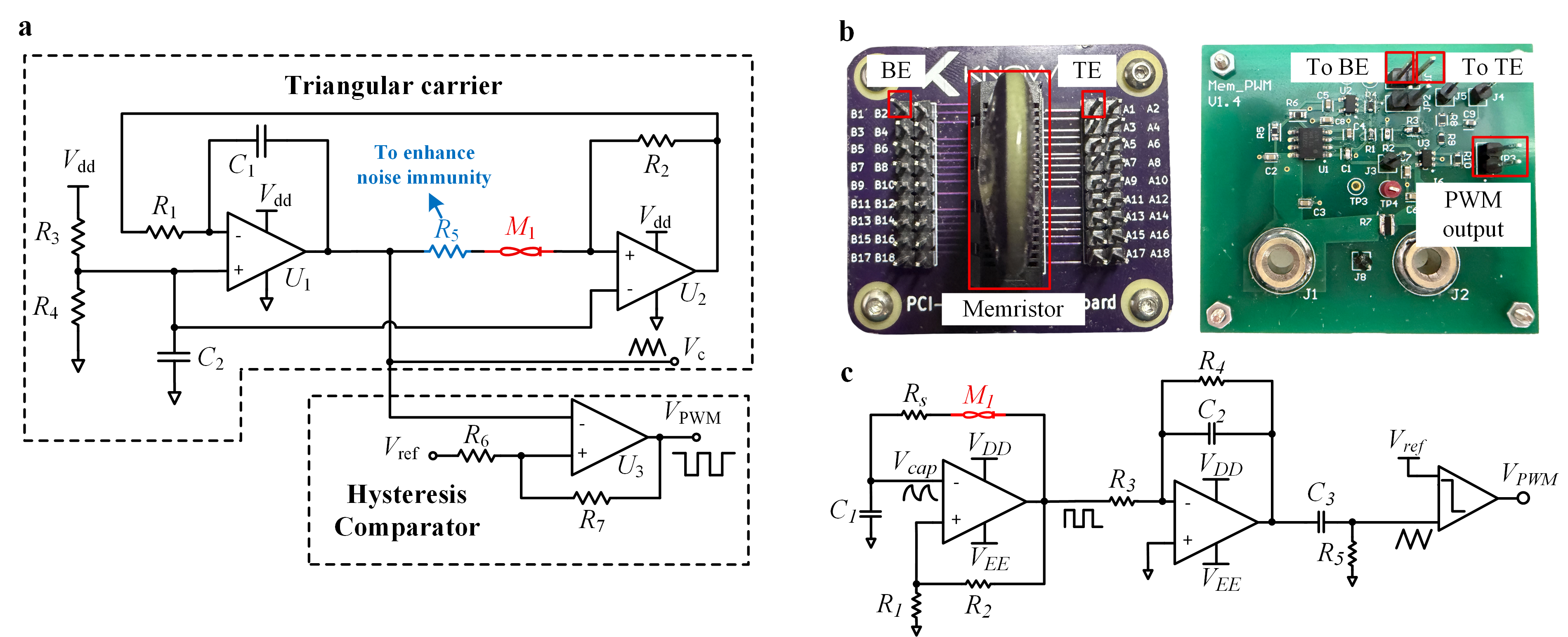}
    \caption{\textbf{Implementation and comparison of the proposed memristor-based PWM circuit.}
    \textbf{a} Schematic of the proposed memristor-based PWM circuit.
    \textbf{b} Prototype of the proposed PWM circuit.
    \textbf{c} Previous memristor-based PWM in \cite{20kHz}.}
    \label{proposed_circuit}
\end{figure}

The proposed memristor-based PWM circuit is shown in Fig.~\ref{proposed_circuit}a, and its prototype is shown in Fig.~\ref{proposed_circuit}b. It consists of two main blocks: a triangular waveform generator that produces the carrier signal and a hysteresis comparator that generates the PWM output. Compared with a conventional analog PWM generator, the fixed resistor is replaced by a memristor, which introduces programmability into the circuit. The triangular carrier generation stage comprises a Schmitt trigger circuit ($U_2$) and an integrator circuit ($U_1$). The Schmitt trigger generates a periodic square-wave signal, which is then applied to the integrator to produce a linearly varying voltage, resulting in a triangular waveform. The frequency of the carrier signal is determined by

\begin{equation}
f = \frac{R_2}{4R_1C_1(R_5+R_{M1})}\label{frequency equation}
\end{equation}
where $R_{M1}$ is the resistance of the memristor. Since the resistance of the memristor can be programmed to different values, different carrier frequencies can be obtained.

After the memristor resistance is programmed to the desired value, any unintended resistance drift during circuit operation can change the oscillation frequency and may lead to incorrect PWM generation. Therefore, the voltage across the memristor must remain below its threshold voltage during normal operation. For the proposed circuit, this condition determines the maximum allowable programmable resistance of the memristor as
\begin{equation}
R_{M1,\max} = \frac{2V_{th}}{V_{dd}}R_2 ,
\label{eq:maxR}
\end{equation}
where $V_{dd}$ is the operational-amplifier supply voltage. This result indicates that increasing the feedback resistor $R_2$ allows a larger allowable memristor resistance range while maintaining the voltage across the memristor below its threshold voltage. Therefore, the proposed topology is suitable for memristor-based PWM generation because the feedback resistor can be used to limit the memristor voltage while improving the oscillation frequency.

For comparison, in the analog PWM circuit reported in \cite{20kHz}, the memristor is placed directly in the feedback loop, as shown in Fig.~\ref{proposed_circuit}c. The oscillation frequency of this circuit is given by
\begin{equation}
f_{old} = \frac{1}{2.2C_1(R_s+R_{M1})}.
\label{eq:frequency_old}
\end{equation}
To keep the voltage across the memristor below its threshold voltage, the maximum allowable memristor resistance in this topology is
\begin{equation}
R_{M1,\max,old} = \frac{2V_{th}}{V_{dd}-2V_{th}}R_s .
\label{eq:maxR_old}
\end{equation}

Equations~(\ref{eq:maxR}) and~(\ref{eq:maxR_old}) show that both topologies require a larger external resistor to extend the allowable memristor resistance range. However, the impact of this resistor on the oscillation frequency is different. Based on Equation (\ref{frequency equation}), larger $R_2$ results in higher frequency. 
In contrast, in the topology of \cite{20kHz}, increasing $R_s$ directly increases the timing resistance in the denominator of~(\ref{eq:frequency_old}), thereby reducing the oscillation frequency. As a result, extending the programmable resistance range in the previous topology comes at the cost of a lower operating frequency, whereas the proposed topology can improve high-frequency operation by increasing $R_2$. This feature makes the proposed circuit more suitable for power converter control, where high-frequency PWM generation is typically required.

Based on~(\ref{frequency equation}) and~(\ref{eq:maxR}), the programmable frequency range of the proposed circuit can be expressed as
\begin{equation}
\frac{R_2V_{dd}}
{4R_1C_1\left(R_5V_{dd}+2V_{th}R_2\right)}
< f <
\frac{R_2}{4R_1C_1\left(R_5+R_{M1,\min}\right)} .
\label{eq:f_range_proposed}
\end{equation}
Similarly, using~(\ref{eq:frequency_old}) and~(\ref{eq:maxR_old}), the programmable frequency range of the previous topology is
\begin{equation}
\frac{V_{dd}-2V_{th}}
{2.2C_1R_sV_{dd}}
< f_{old} <
\frac{1}{2.2C_1\left(R_s+R_{M1,\min}\right)} .
\label{eq:f_range_old}
\end{equation}

Equation~(\ref{eq:f_range_old}) shows that the achievable frequency in the previous topology is inversely proportional to the series resistance $R_s$. Therefore, increasing $R_s$ to enlarge the allowable memristor resistance range reduces both the minimum and maximum operating frequencies. In contrast, in the proposed circuit, the feedback resistor $R_2$ can be increased to extend the allowable memristor resistance range while also supporting higher-frequency oscillation. This feature provides a wider and more practical programmable frequency range for memristor-based PWM control. Moreover, the previous design requires both positive and negative voltage rails, whereas the proposed design requires only a positive voltage rail, thereby improving circuit simplicity.

Another key consideration for ensuring proper circuit operation is that the triangular carrier signal must have sufficient amplitude to provide sufficient noise immunity. Without the series resistor $R_5$, the maximum peak-to-peak amplitude of the carrier signal with $R_{M1,max}$ is 
\begin{equation}
V_{c,pp,\max} = 2V_{th}
\label{Vcmax1}
\end{equation}
Since the threshold voltage of currently available commercial memristors is relatively low and a safety margin must be maintained during operation, the maximum carrier amplitude would be limited to less than $200~\text{mV}$, making it highly sensitive to noise. Therefore, the series resistor $R_5$ is introduced to increase the peak-to-peak amplitude of the carrier signal, which becomes
\begin{equation}
V_{c,pp,\max} = \frac{R_5}{R_2}V_{dd} + 2V_{th}
\label{Vcmax2}
\end{equation}
With an appropriate choice of $R_5$ and the use of a hysteresis comparator in the final PWM stage, the circuit achieves improved noise immunity. 

With the carrier signal \( V_{c} \), the PWM output is generated by comparing it with a reference signal using a hysteresis comparator. Introducing hysteresis in the comparator improves the robustness and noise immunity of the PWM generation process. In the absence of hysteresis, small disturbances or noise on the carrier signal $V_c$ or reference signal $V_{ref}$ may cause unintended rapid switching at the comparator output, leading to false triggering and instability in \( V_{\text{PWM}} \). To mitigate this issue, a pair of resistors \( R_6 \) and \( R_7 \) are used to implement hysteresis, resulting in two distinct switching thresholds: the upper threshold \( V_{\text{th}+} \) for the low-to-high transition and the lower threshold \( V_{\text{th}-} \) for the high-to-low transition. This configuration suppresses false switching and ensures stable and well-defined PWM transitions. Such a hysteresis-based comparator is especially beneficial in applications requiring accurate timing and reliable PWM signal generation. The threshold voltages \( V_{\text{th}+} \) and \( V_{\text{th}-} \) of the comparator are given by

\begin{equation}
\begin{aligned}
V_{\text{th}+} &= \frac{R_6}{R_6 + R_7} V_{OH} + \frac{R_7}{R_6 + R_7} V_{ref} \\
V_{\text{th}-} &= \frac{R_6}{R_6 + R_7} V_{OL} + \frac{R_7}{R_6 + R_7} V_{ref}
\end{aligned}
\end{equation}
where $V_{OH}$ is the high output voltage of the comparator and $V_{OL}$ is the low output voltage of the comparator. 



\subsection*{Programmable PWM generation}
In the experiment, a Knowm M+SDC Cr-type memristor chip is used with a typical threshold voltage of 0.33 V \cite{Knowm_datasheet}. The memristor is programmed to the desired resistance value using the programming setup and procedures shown in Fig.~\ref{Mem_prog}, and then integrated into the proposed circuit for PWM generation.

\begin{figure}[htbp]
    \centering
    \includegraphics[
        width=0.95\textwidth,
        ]{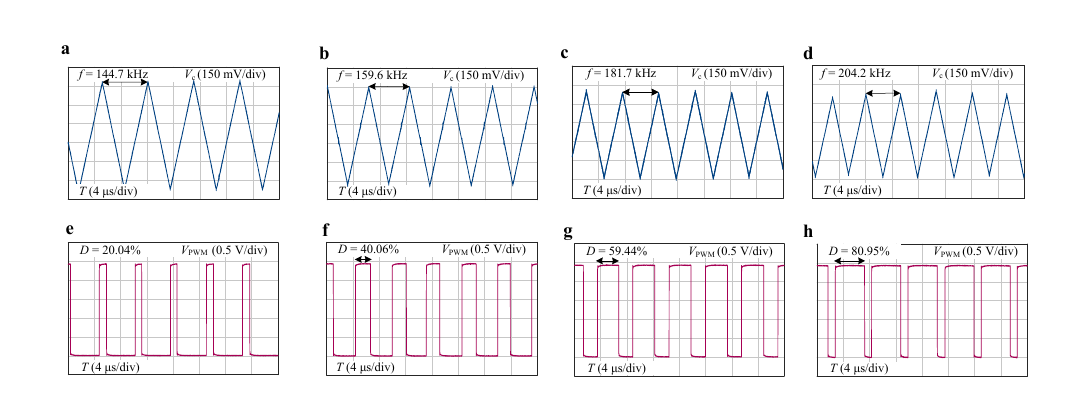}
    \caption{\textbf{Experimental validation of programmable PWM generation.}
    \textbf{a--d} Measured triangular-carrier waveforms at programmed switching frequencies of 144.7, 159.6, 181.7 and 204.2~kHz, respectively.
    \textbf{e--h} Measured PWM output waveforms at 181.7~kHz with duty cycles of 20.04\%, 40.06\%, 59.44\% and 80.95\%, respectively.}
    \label{PWM_results}
\end{figure}

The experimental results of the output switching frequencies for different programmed resistance values are summarized in Table~\ref{tab:freq_compare}. The waveforms of the carrier signals corresponding to different memristor resistance values are shown in Fig.~\ref{PWM_results}a--d. The output PWM waveforms at 181.7~kHz with duty cycles of 0.2, 0.4, 0.6, and 0.8 are presented in Fig.~\ref{PWM_results}e--h.

The experimental results demonstrate the programmability of the proposed PWM circuit through adjustment of the memristor resistance. By programming the memristor to different resistance values, the frequency of the PWM output can be effectively tuned over a wide range. A discrepancy is observed between the calculated switching frequencies and the experimentally measured values. This deviation is attributed primarily to nonideal effects in the hardware implementation, including parasitic capacitances associated with the prototype circuit. These parasitic elements may affect the effective impedance in Equation~(\ref{frequency equation}), resulting in a deviation in the operating frequency. 
\begin{table}[htbp]
\centering
\caption{Comparison of calculated and measured switching frequencies for different memristor resistance values.}
\label{tab:freq_compare}
\small
\setlength{\tabcolsep}{12pt}
\renewcommand{\arraystretch}{0.95}
\begin{tabular}{cccc}
\toprule
$R_{\mathrm{M1}}$ (k$\Omega$) &
Calculated $f$ (kHz) &
Measured $f$ (kHz) &
Difference (\%) \\
\midrule
72.803 & 144.67 & 144.7 & $-0.02$ \\
49.745 & 166.95 & 159.6 & 4.40 \\
21.374 & 205.97 & 181.7 & 11.79 \\
6.983  & 233.68 & 204.2 & 12.62 \\
\bottomrule
\end{tabular}
\end{table}

\subsection*{Buck-converter validation and PWM-generation power}
\begin{figure}[htbp]
    \centering
    \includegraphics[
        width=0.9\textwidth,
        ]{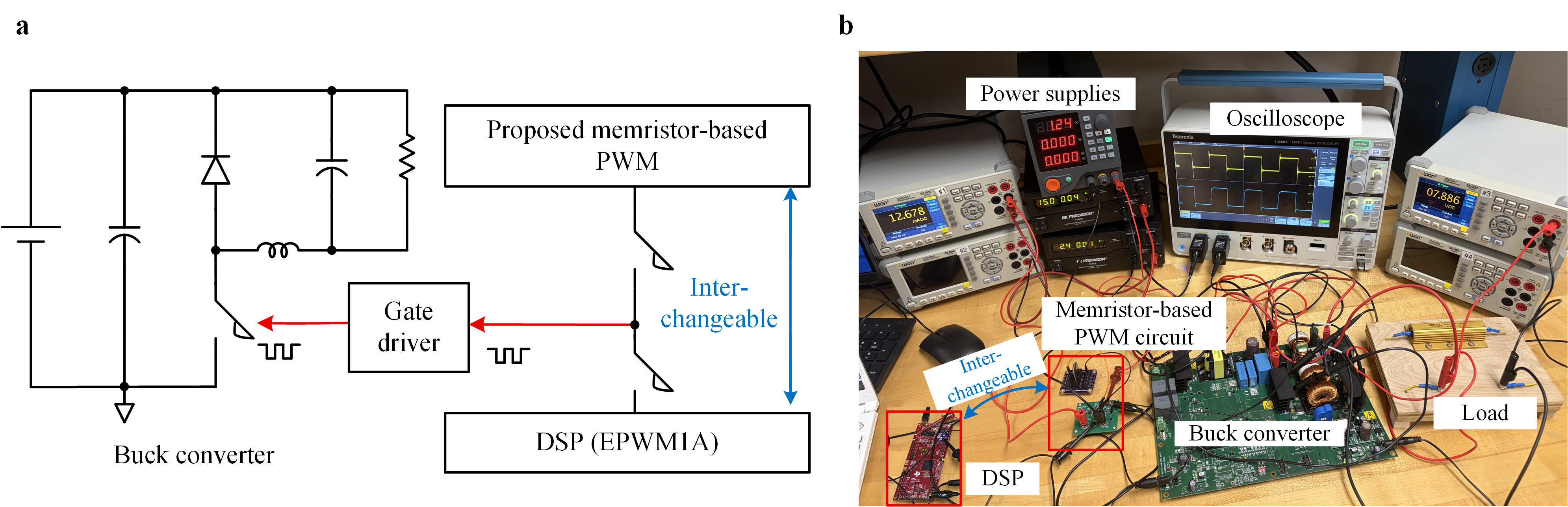}
    \caption{\textbf{Experimental configuration for buck-converter validation and PWM-generation power comparison.}
    \textbf{a}, Schematic of the buck-converter test platform. 
    \textbf{b}, Photograph of the experimental setup.}
    \label{buck_setup}
\end{figure}

To evaluate the power efficiency of the proposed memristor-based PWM circuit, its power consumption is compared with that of a widely used DSP controller, the TI LAUNCHXL-F28379D. In both cases, a single PWM signal is generated to control a buck converter implemented using one half-bridge of the TIDM-SOLARUINV micro solar inverter as shown in Fig.~\ref{buck_setup}a. This ensures a fair comparison by limiting both controllers to identical control functionality. Due to the operational constraints of the semiconductor in the buck converter, the switching frequency for this comparison is set to 100~kHz.

For the DSP-based implementation, only the EPWM1A module is enabled, while all other peripherals and control functions are disabled. For the memristor-based implementation, only the PWM generation circuit is powered. The power stage, gate driver, and auxiliary circuits are excluded from the power consumption measurement in both cases. Both controllers are powered using an external bench power supply. The supply voltage and input current are measured directly using calibrated digital multimeters, and the power consumption is calculated from the measured voltage and current values. This approach ensures consistent measurement conditions for both implementations and avoids additional uncertainty introduced by onboard regulators or USB power sources. A conventional analog PWM circuit is also tested by replacing the memristor with a resistor on the board. The experiment setup is shown in Fig.~\ref{buck_setup}b. The input voltage of the buck converter is 15 V.

\begin{table}[t]
\centering
\caption{Buck-converter output voltages at different duty cycles.}
\label{tab:buck_compare}
\small
\setlength{\tabcolsep}{14pt}
\renewcommand{\arraystretch}{0.95}
\begin{tabular}{lcc}
\toprule
\textbf{Controller} &
\textbf{Duty cycle (\%)} &
\textbf{$V_{\mathrm{out}}$ (V)} \\
\midrule
Memristor-based PWM
& 23.54 & 4.503 \\
& 49.57 & 7.886 \\
& 74.74 & 11.468 \\
\midrule
TI F28379D
& 24.51 & 4.598 \\
& 49.31 & 7.818 \\
& 74.54 & 11.439 \\
\bottomrule
\end{tabular}
\end{table}

\begin{table}[t]
\centering
\caption{Power consumption comparison of the evaluated PWM controllers.}
\label{tab:power_compare}
\small
\setlength{\tabcolsep}{12pt}
\renewcommand{\arraystretch}{0.95}
\begin{tabular}{lcc}
\toprule
\textbf{Controller} &
\textbf{$P_{\mathrm{in}}$ (mW)} &
\textbf{Programmable} \\
\midrule
Proposed memristor-based PWM & \textbf{30.36} & \textbf{Yes} \\
DSP (EPWM1A only)          & 338.38 & Yes \\
Conventional analog PWM    & 30.43  & No \\
\bottomrule
\end{tabular}
\end{table}

The buck converter output voltages under different duty cycle using the memristor-based PWM circuit and the DSP-based PWM implementation are shown in Table~\ref{tab:buck_compare}. The results indicate that the proposed circuit is capable of generating PWM signals with comparable quality to those generated by the DSP. The measured power-consumption results are summarized in Table~\ref{tab:power_compare}. The proposed memristor-based PWM circuit exhibits significantly lower power consumption than the DSP-based implementation, achieving a reduction of approximately 92\%, which is similar to the conventional analog PWM circuit. This improvement is attributed to the analog operation of the proposed circuit, which avoids the Von Neumann architecture and its associated clocking, memory access, and digital-logic overhead in DSP-based controllers. These results demonstrate the strong potential of memristor-based PWM generation for ultra-low-power control in power electronics applications.

\subsection*{Comparison with prior memristor-based PWM approaches}
To further quantify the advantage of the proposed memristor-based PWM circuit, a comparison study is conducted with previous memristor-based PWM topologies. The comparison focuses on two key aspects: the allowable programmable resistance range of the memristor and the corresponding achievable programmable frequency range. A wider allowable resistance range provides more usable programmed states and improves the practicality of frequency tuning. In addition, a wider achievable frequency range gives more flexibility for power converter control. 

The minimum programmable resistance of the memristor in the prior circuit \cite{20kHz} is determine by (\ref{eq:preferred_range_general}) and (\ref{eq:preferred_range_default}), and the maximum allowable programmable resistance is determined by (\ref{eq:maxR_old}). 
To provide a fair comparison at an operating frequency of approximately
$200~\mathrm{kHz}$, the required series resistance of the previous design
is calculated using (11). With
$R_{\mathrm{mem,min}}=1~\mathrm{k}\Omega$ and
$C=100~\mathrm{pF}$, the required series resistance is
$R_s \approx 21.7~\mathrm{k}\Omega$.
With this series resistance, the maximum allowable memristance is
determined by
\begin{equation}
R_{\mathrm{mem,max}}
=
\frac{2V_{\mathrm{th}}}
{V_{\mathrm{dd}}-2V_{\mathrm{th}}}R_s
\approx 7.97~\mathrm{k}\Omega.
\label{eq:old_rmem_max}
\end{equation}
Therefore, the allowable memristance range of the previous design is
\begin{equation}
1~\mathrm{k}\Omega
\leq R_{\mathrm{mem,old}}
\leq 7.97~\mathrm{k}\Omega.
\end{equation}
Substituting these resistance limits into (11) gives the corresponding theoretical
programmable frequency range as
\begin{equation}
156.9~\mathrm{kHz}
\leq f_{\mathrm{old}}
\leq 206.6~\mathrm{kHz}.
\end{equation}
The theoretical memristor-programmable range and achievable frequency of the proposed circuit can be obtained from (\ref{eq:maxR}) and (\ref{eq:f_range_proposed}). The comparison is shown in the Table~\ref{tab:range_comparison}.

\begin{figure}[htbp]
    \centering
    \includegraphics[
        width=0.62\textwidth,
        height=0.48\textheight,
        keepaspectratio
    ]{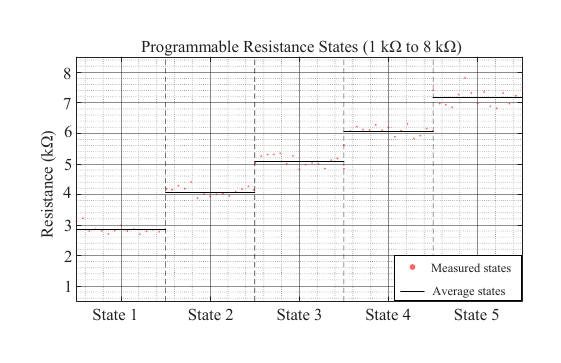}
    \vspace{-15pt}
    \caption{\textbf{Experimentally programmed low-resistance states of the KNOWM Cr-type memristor.}
    Five distinguishable resistance states were programmed within the range from approximately 1 to 8~k$\Omega$. Red markers indicate the measured resistance values obtained from repeated programming trials, and black horizontal lines indicate the average resistance of each programmed state.}
    \label{1kto8k}
\end{figure}

\begin{table}[t]
\centering
\caption{Comparison of theoretical programmable resistance and frequency ranges.}
\label{tab:range_comparison}
\small
\setlength{\tabcolsep}{10pt}
\renewcommand{\arraystretch}{0.95}
\begin{tabular}{lccc}
\toprule
\textbf{Metric} &
\textbf{Prior work~\cite{20kHz}} &
\textbf{Proposed} &
\textbf{Increase} \\
\midrule
Memristance range
& 1--7.966 k$\Omega$
& 1--129.25 k$\Omega$
& 18.41$\times$ \\
Frequency bandwidth
& 49.689 kHz
& 138.47 kHz
& 2.79$\times$ \\
\bottomrule
\end{tabular}
\end{table}
As summarized in Table~\ref{tab:range_comparison}, the proposed topology extends the allowable memristance range by $18.41\times$ and increases the programmable frequency bandwidth by $2.79\times$ compared with the previous topology.
Moreover, the achievable resistance states of the Knowm memristor between 1 k$\Omega$ and 8 k$\Omega$ are shown in Fig. \ref{1kto8k}.
Although the previous topology can still realize at least five distinct nominal frequency states within its allowable resistance range, the narrower resistance range leads to smaller separation between adjacent programmed states. As a result, resistance-programming variations can produce larger frequency deviations and may cause overlap between neighboring frequency intervals. In contrast, the proposed topology provides a much wider programmable resistance range, allowing the resistance states to be spaced farther apart and thereby improving the distinguishability and robustness of the corresponding frequency states.

\begin{figure}[t]
    \centering
    \includegraphics[
        width=\textwidth,
        height=0.62\textheight,
        keepaspectratio
    ]{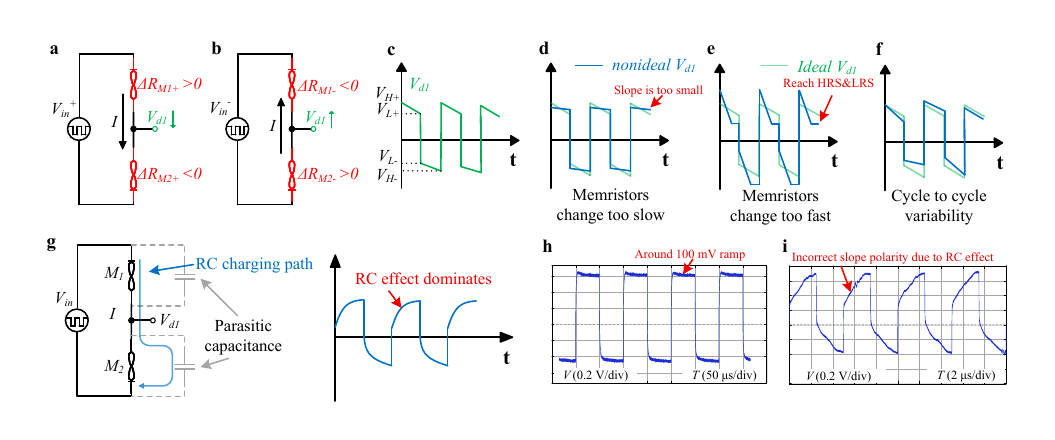}
    \caption{\textbf{Operating principle and nonideal limitations of the previous two-memristor carrier-generation topology.}
    \textbf{a} Resistance changes of memristors $M_1$ and $M_2$ during the positive half-cycle of the applied voltage.
    \textbf{b} Resistance changes during the negative half-cycle, with the resistance-update directions reversed.
    \textbf{c} Ideal divider-voltage waveform obtained under matched and repeatable resistance changes.
    \textbf{d} Reduced carrier slope when the memristor resistance changes too slowly.
    \textbf{e} Premature saturation when the memristors reach their high- and low-resistance states before the end of the half-cycle.
    \textbf{f} Cycle-to-cycle carrier variation caused by stochastic resistance updates.
    \textbf{g} Parasitic-capacitance-induced charging path in the two-memristor voltage divider and the resulting RC-dominated waveform at high frequency.
    \textbf{h} Measured divider-voltage waveform showing a carrier ramp of approximately 100~mV at 10 kHz.
    \textbf{i} Measured high-frequency waveform showing an incorrect slope polarity when the parasitic RC response dominates at 200 kHz.}
    \label{Topology10Hz}
\end{figure}

Another factor preventing prior designs from achieving reliable PWM operation is the inherent stochasticity and non-ideal behavior of practical memristors. The circuit in \cite{10HzPWM} connects two memristors in series
with opposite polarities. The memristor pair is driven by a square-wave
input, $V_{\mathrm{in}}$, as shown in Fig.~\ref{Topology10Hz}. When
$V_{\mathrm{in}}$ is positive, as illustrated in
Fig.~\ref{Topology10Hz}a, the resistance of $M_1$ increases while that
of $M_2$ decreases. Owing to the memristor dynamics, the divider output
voltage $V_{d1}$ exhibits a decreasing slope and can therefore be used as
a sawtooth carrier signal. A similar process occurs when
$V_{\mathrm{in}}$ is negative, as shown in Fig.~\ref{Topology10Hz}b.
As a result, the voltage-divider output $V_{d1}$ forms a sawtooth-like
waveform that serves as the carrier signal, as shown in
Fig.~\ref{Topology10Hz}c. To ensure proper circuit operation, the two
memristors must return to their initial resistance states at the
beginning of each cycle so that the output can satisfy:
\begin{equation}
\begin{gathered}
\left|\Delta R_{M1+}\right|
=
\left|\Delta R_{M2+}\right|
=
\left|\Delta R_{M1-}\right|
=
\left|\Delta R_{M2-}\right|,
\\[2pt]
\left|V_{H+}\right|
=
\left|V_{H-}\right|,
\qquad
\left|V_{L+}\right|
=
\left|V_{L-}\right|,
\\[2pt]
\left|V_{H+}\right|-\left|V_{L+}\right|
=
\left|V_{H-}\right|-\left|V_{L-}\right|.
\end{gathered}
\label{eq:memristor_reset_condition}
\end{equation}

However, this behavior is obtained using ideal memristor models, whereas the nonideal characteristics of practical memristors may cause circuit malfunction. Different memristors exhibit different sensitivities to the applied voltage pulses. If a memristor is insufficiently sensitive, its resistance may change only slightly, as shown in Fig.~\ref{Topology10Hz}d. Consequently, the divider voltage $V_{d1}$ cannot develop the required ramp within one half-cycle, resulting in a reduced carrier amplitude and degraded noise immunity. Although an additional amplification stage could be used to increase the ramp amplitude, it would increase circuit complexity and could also amplify noise.
Conversely, if the memristor is overly sensitive to the applied pulses, its resistance may change too rapidly, causing the devices to reach their HRS and LRS before the end of the half-cycle. As shown in Fig.~\ref{Topology10Hz}e, $V_{d1}$ then saturates near the corresponding voltage limits and produces a flattened segment instead of a continuous ramp, rendering the carrier signal unusable.
Furthermore, device-to-device variation and cycle-to-cycle stochastic resistance changes may prevent the two memristors from returning to the same initial resistance states at the beginning of each cycle. Consequently, the peak value, slope, and amplitude of $V_{d1}$ may vary from cycle to cycle, as illustrated in Fig.~\ref{Topology10Hz}f. Such variability undermines the intended operation of the circuit and may lead to an unstable or inaccurate PWM signal. In addition, package parasitic capacitance may distort the output waveform at high operating frequencies. If the input switches before the parasitic RC network has fully charged or discharged, an incorrect output waveform may be produced, as shown in Fig.~\ref{Topology10Hz}g.
Some experiments using the topology in Fig. \ref{Topology10Hz}a were also conducted with Knowm memristors. The square-wave input was generated using a function generator with an amplitude of $V_{in}=\pm 2$~V and was applied at frequencies of 10~kHz and 200~kHz.

At $10~\mathrm{kHz}$, as shown in Fig.~\ref{Topology10Hz}h, the divider voltage $V_{d1}$ is nearly a square wave, with a ramp of only approximately $100~\mathrm{mV}$ produced during each half-cycle. This indicates that the resistance changes only slightly during each half-cycle, resulting in an insufficient carrier amplitude and poor noise immunity, which is consistent with the nonideal behavior shown in Fig.~\ref{Topology10Hz}d.
When the operating frequency is increased to $200~\mathrm{kHz}$, as shown in Fig.~\ref{Topology10Hz}i, the waveform becomes strongly nonlinear and exhibits a slope opposite to the desired direction. At this frequency, the circuit response is dominated by the charging and discharging of parasitic capacitances associated with the device packaging and circuit connections, rather than by the intended memristor-state dynamics. This behavior is consistent with Fig.~\ref{Topology10Hz}g. Consequently, the topology fails to generate a stable and usable sawtooth carrier under the tested practical operating conditions.

\section*{Discussion}
Theoretically, the memristor can be integrated into a conventional analog PWM circuit, enabling the generation of PWM signals with different frequencies. However, the characteristics of the currently available commercial memristor exhibit several limitations on the circuit performance. In particular, the relatively low threshold voltage limits the allowable voltage across the memristor during normal operation, which restricts the achievable programmable range of the frequency and the carrier-signal amplitude. A higher threshold voltage would allow a larger carrier amplitude while maintaining the memristor below its switching threshold, thereby improving noise immunity and expanding the usable resistance range.

Nevertheless, the threshold voltage cannot be increased without limit. In power-electronics control circuits, the analog supply voltage is typically limited to $3.3~\mathrm{V}$ or $5~\mathrm{V}$. Therefore, if the memristor threshold voltage is too high, the available supply voltage may not provide sufficient voltage headroom to program the resistance state. This requirement leads to a preferred threshold-voltage range: the threshold voltage should be high enough to prevent unintended resistance drift during PWM operation, but low enough to allow practical programming using the available analog supply.

\begin{figure}[t]
    \vspace{-15pt}
    \centering
    \includegraphics[
        width=0.62\textwidth,
        height=0.48\textheight,
        keepaspectratio
    ]{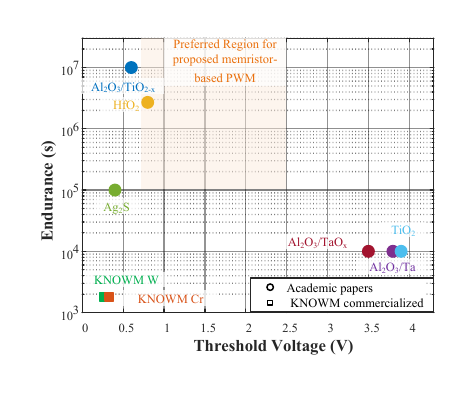}
    \vspace{-15pt}
    \caption{\textbf{Comparison of memristor threshold voltage and endurance for the proposed memristor-based PWM circuit.}}
    \label{compare_mem}
\end{figure}

Various memristor devices with different materials, threshold voltages, and endurance characteristics have been reported in the literature~\cite{HfO2_1,HfO2_2,PDcontrol,Ag2S,TiO2_1,TiO2_2,Al2O3/TaOX,Al2O3/TiO2-x}. A comparison between representative academic devices and the commercial Knowm memristors is shown in Fig.~\ref{compare_mem}. For the proposed memristor-based PWM circuit, devices with higher endurance and moderate threshold voltage are preferred. Higher endurance time supports longer operation without unintended self-changes in the memristor state, while a moderate threshold voltage provides a suitable tradeoff among operational stability, noise immunity, programmable range, and low-voltage programmability.

\section*{Methods}
\subsection*{Memristor characterization and resistance programming}
Both commercial Knowm M+SDC tungsten-type and chromium-type memristors were used in the characterization and programming experiments. The devices were mounted on a Knowm Discovery V2 board and operated using the associated software interface. The equivalent circuit of the memristor connected to the board is shown in Fig.~\ref{Mem_prog}b.
The current–voltage hysteresis characteristics shown in Fig.~\ref{Mem_fundamental}c were measured using the Hysteresis experiment mode. A sinusoidal voltage was applied to the device, and the resulting current–voltage trajectory was recorded from the I–V waveform window. The excitation amplitude and frequency were set to 0.8 V and 10 Hz. Each hysteresis curve represents 1 cycle.
Resistance programming was performed using the Square experiment mode. The closed-loop programming procedure shown in Fig.~\ref{Mem_prog}d was applied, beginning with a programming pulse of 0.8 V and 50 $\mu$s. The resistance was measured after each pulse using a read voltage of 0.1 V. The real-time resistance was monitored through the software interface.

\subsection*{PWM circuit implementation}

The PWM prototype was assembled on a printed circuit board, as shown in Fig.~\ref{proposed_circuit}b. Most of the analog circuit was implemented directly on the main board. The memristor chip was mounted on a Knowm breakout board and connected to the PWM board through jumper wires, allowing different devices to be evaluated using the same circuit.
The triangular-carrier generator consisted of an integrator implemented using an OPA2323 operational amplifier and a Schmitt trigger implemented using a TLV3601 high-speed comparator. A second TLV3601 was configured as the hysteresis comparator used to generate the PWM output by comparing the triangular carrier with the reference voltage. The carrier and PWM waveforms were measured using a Tektronix MDO34 oscilloscope with passive probes. The component values and part numbers used in the prototype are summarized in Table~\ref{tab:circuit_parameters}.

\begin{table}[htbp]
\centering
\caption{Component values and part numbers of the proposed PWM circuit.}
\label{tab:circuit_parameters}
\small
\setlength{\tabcolsep}{18pt}
\renewcommand{\arraystretch}{0.95}
\begin{tabular}{lc}
\toprule
\textbf{Component} & \textbf{Value or part number} \\
\midrule
$R_1$             & 47 k$\Omega$ \\
$R_2$             & 470 k$\Omega$ \\
$R_3$, $R_4$      & 100 $\Omega$ \\
$R_5$             & 100 k$\Omega$ \\
$R_6$             & 10 k$\Omega$ \\
$R_7$             & 220 k$\Omega$ \\
$C_1$             & 100 pF \\
$U_1$             & OPA2323 \\
$U_2$, $U_3$      & TLV3601 \\
$V_{\mathrm{dd}}$ & 2.4 V \\
\bottomrule
\end{tabular}
\end{table}

\subsection*{Buck-converter validation and power consumption measurements}
One half-bridge leg of a TI TIDM-SOLARUINV microinverter platform was configured to operate as a buck converter, with the upper switch used as the active device. The experimental configuration is shown in Fig.~\ref{buck_setup}. The proposed memristor-based PWM circuit and a TI LAUNCHXL-F28379D development board were used interchangeably to generate
a single PWM signal for the same gate driver, thereby maintaining identical converter operating conditions for both controllers.
For the power-consumption comparison, the two controllers were powered separately from a B\&K Precision 1685B
 bench power supply. In the DSP-based implementation, only the EPWM1A module was enabled, while the remaining peripherals and control functions were disabled. In the memristor-based implementation, only the PWM-generation circuit was powered. The gate driver, converter power stage, and auxiliary circuits were excluded from both measurements.
The supply voltage and input current of each controller were measured using OWON XDM2041 digital multimeters. The controller input power was calculated as the product of the measured supply voltage and current.

\section*{Data availability}
The data supporting the findings of this study are available from the corresponding author upon request. 

\section*{Code availability}
Programming and analysis scripts used to process the experimental data are available from the corresponding author upon request.

\section*{Acknowledgments}
This work was supported by the U.S. National Science Foundation under
Award No.~2339806.

\section*{Author contributions}
F.W. conceived the study, conducted the experiments, analyzed the data, and wrote the paper.
X.L. conducted the experiments and edited the manuscript.
Y.L. supervised the research, acquired funding, and reviewed and edited
the manuscript. All authors reviewed and approved the final manuscript.

\section*{Competing interests}
The authors declare no competing interests.

\bibliographystyle{naturemag}
\bibliography{reference}
\end{document}